\documentclass[aps,prd,showpacs]{revtex4}
\usepackage{epsfig,amsmath,amsfonts}
\usepackage{dsfont,graphicx}
\usepackage{graphicx}
\newcommand{\bs}{\begin{subequations}}
\newcommand{\es}{\end{subequations}}
\newcommand{\adb}{\allowdisplaybreaks }
\newcommand{\os}[2]{\overset{#1}{#2}{}}
\usepackage{color}
 \definecolor{darkgreen}{rgb}{0,.5,0}
 \usepackage[unicode,colorlinks,filecolor=blue,citecolor=darkgreen]{hyperref}
\begin{document}
\title{Bremsstrahlung in wormhole spacetime with infinitely short throat}
\author{Nail Khusnutdinov\footnote{e-mail: nail.khusnutdinov@gmail.com}}
\affiliation{Institute of Physics, Kazan Federal University, Kremlevskaya 18, Kazan, 420008, Russia \\
and\\ Department of Physics, University of South Florida, Tampa, Florida 33620, USA}

\begin{abstract}
We consider the total energy loss and spectral density of uniformly moving electrically charged particles in the spacetime of a wormhole with an infinitely short throat. We show that the total energy loss $\mathcal{E} \sim e^2v\gamma a^2/b^3$, where $\gamma$ is relativistic factor, $a$ is the radius of the wormhole's throat and $b$ is the impact factor. The spectrum of the energy for particles radially moving through the wormhole's throat  $\mathcal{E} \sim e^2v\gamma/a$. The spectral density of the total energy has a maximum at frequency $\omega_m \sim v\gamma/b$ and at $\omega_m \sim v\gamma /a$ for radial motion.     
\end{abstract}  


\maketitle 

\section{Introduction}

Wormholes are topological handles that join different regions of the Universe or different universes. A  wormhole has to violate energy conditions and thus a source of the wormhole geometry should be an exotic matter. By now many different approaches were suggested to avoid this problem and find a self-consistent solution for the wormhole space-time.  Vacuum fluctuations of quantum fields may serve this purpose   \cite{Krasnikov:2000:tw,Khusnutdinov:2002:Gsews,Khusnutdinov:2003:Sw,Garattini:2005:Sstw}; in the framework of multidimensional gravity the additional dimensions contribute to the $3+1$ dimensional Einstein equations \cite{Shiromizu:2000:TEebw} used by Bronnikov \cite{Bronnikov:2003:Pwbw} to find possible metrics of wormholes. The cosmic phantom energy \cite{Sushkov:2005:Wsbape,Lobo:2005:Petw}  or reverse sign of the kinetic term \cite{ArmendarizPicon:2002:cstLwcgr} were used to solve this problem. An introduction in wormhole physics may be found in the Visser book \cite{Visser:1995:LWfEtH} and the Lobo review   \cite{Lobo:2008:CaQGR}. Some astronomical aspects of the astrophysics of wormholes were discussed in Refs. \cite{Kardashev:2007:AW,Novikov:2007:mUaw,Shatskii:2009:Tsoauastaw}. 

A charged particle in the spacetime of a wormhole is attracted to the wormhole's throat by an additional gravitationally induced self-interaction force \cite{Khusnutdinov:2007:Scpwst,Krasnikov:2008:Eipcw,Bezerra:2009:Sspwst,Khusnutdinov:2010:Spcsmw,Boisseau:2013:Eiaswr}. This force is a manifestation of the nonlocal essence
of the electromagnetic field. A charged particle will be attracted by wormholes, even though it is at rest.  From an astrophysical point of view this means that a wormhole's throat should be surrounded by a cloud of cosmological particles.    

It is well known that in flat Minkowsky spacetime uniformly moving charged particle does not produce  electromagnetic radiation.  In the framework of quantum electrodynamics, the bremsstrahlung process corresponds to the emission of radiation by a charged particle when it changes its momentum in collisions with obstacles such as other particles or when it is accelerated due to the presence of electromagnetic fields. In curved spacetime the situation is quite different -- a uniformly moving charged particle produces radiation. Uniform motion in curved spacetime is motion along the geodesic line.  Aliev \& Galtsov in Ref. \cite{Aliev:1989:Garsgcs} were the first to establish this effect in the context of cosmic strings. The spacetime of the cosmic string is everywhere flat except at its origin with infinite curvature at the core and a uniformly moving particle moves on a straight line. They calculated the total energy loss and its spectral density for particles moving on these lines. This classical result was reviewed and recovered in the framework of quantum field theory in Refs.     \cite{Audretsch:1991:Cbcss,Audretsch:1991:QftpncsTpl,Skarzhinsky:1994:Qegfcs}. In Refs. \cite{Audretsch:1991:Cbcss,Audretsch:1991:QftpncsTpl} the bremsstrahlung was considered for a simple scalar model in which the electron-positron quantum field models by charged scalar field and photon field considered as uncharged scalar field.  In Ref. \cite{Skarzhinsky:1994:Qegfcs} this process was considered in the framework of quantum electrodynamics. The same approach was applied for a point-like global monopole space-time in Ref. \cite{Bezerra:2002:Bgfgm}. 

In the present paper we consider an electrically charged and uniformly moving particle in the wormhole spacetime and calculate its total energy loss and spectral density of the energy. All trajectories may be roughly divided into two classes. The first class contains the world lines of particles which move through the wormhole throat. These particle disappear from the point of view of an observer situated in one part of the wormhole spacetime. The rest of the trajectories belong to the second class. We consider radiation for both kind of trajectories. 

There are many different metrics of wormholes, a review of some may be found in Ref.  \cite{Lobo:2008:CaQGR}. We consider the simplest spherically symmetric wormhole with  metric 
\begin{equation}
ds^2 = -dt^2 + d\rho^2 + r^2(\rho) d\Omega,\label{eq:metric}
\end{equation}
considered by Bronnikov \cite{Bronnikov:1973:Stasc} and Ellis  \cite{Ellis:1973:Eftdpmgr}, where $d\Omega$ is the metric of the unit 2D sphere and $t,\rho \in \mathds{R}$.  The spacetime is divided into two parts in which $\rho >0$ and $\rho <0$. The function $r(\rho)$ describes the profile of the wormhole's throat. The radius of the throat, $a$, is defined as the minimum of this function at the point $\rho = 0$: $r(0) = a$. Non-zero components of the Ricci tensor and scalar curvature read
\begin{equation}
R^\rho_\rho = -2 \frac{r''}{r},\ R^\theta_\theta = R^\varphi_\varphi = -\frac{r r''+ r'^2 - 1}{r^2}, R = -\frac{2 \left(2 r r''+r'^2-1\right)}{r^2}.
\end{equation}
The metric with profile $r(\rho) = \sqrt{\rho^2 + a^2}$ is commonly called "drainhole"\/ \cite{Ellis:1973:Eftdpmgr}. In the present paper we consider the simplest model of the wormhole  with infinitely short throat described by profile $r(\rho ) = a +|\rho|$  which was suggested in Ref. \cite{Khusnutdinov:2002:Gsews}. This spacetime is two copies of Minkowsky spacetimes which is glued on the sphere with radius of the throat $a$. The spacetime is everywhere flat except this sphere where tensor Ricci has singular form  
\begin{equation}
R^\rho_\rho = - \frac{4}{a} \delta (\rho),\ R^\theta_\theta = R^\varphi_\varphi = - \frac{2}{a} \delta (\rho),\ R = - \frac{8}{a} \delta (\rho),
\end{equation}
because $r' = \textrm{sgn} \rho, r'' = 2 \delta(\rho)$. Here $\delta (x)$ is delta function. The "drainhole"\/ with profile $r(\rho ) =\sqrt{a^2 +\rho^2}$ has regular Ricci tensor  
\begin{equation}
R^\rho_\rho = R = - \frac{2a^2}{(a^2+\rho^2)^2}.
\end{equation}

The organization of this paper is as follows. In Sec. \ref{Sec:EMF} we consider the Maxwell equations and find vector Green functions we need. In Sec. \ref{Sec:Brems} we develop a general approach to calculate of the total energy loss and its spectral density. We apply this approach to a particle which uniformly moves with nonzero impact parameter as well as for radial motion of the particle and analyze the expression obtained for non-relativistic and ultra-relativistic cases.  We conclude with a discussion of the results in Sec. \ref{Sec:DC}.

\section{Electromagnetic field of charged particle}\label{Sec:EMF}

The Maxwell equations in the Lorentz gauge $A^\nu_{\ ;\nu} = 0$ read 
\begin{equation}
\square A_\mu - R_\mu^\nu A_\nu = - 4\pi J_\mu,
\end{equation}
where $\square = g^{\mu\nu}\nabla_\mu\nabla_\nu$, and the electrical current density has the following form  
\begin{equation}
J^\mu = \frac{e}{\sqrt{-g}} \int d\tau \delta^{(4)} (x - x(\tau))u^\mu (\tau) = \frac{e}{\sqrt{-g}}\frac{u^\mu (\tau^*)}{u^t (\tau^*)}\delta^{(3)} (\vec{x} - \vec{x}(\tau^*)).\label{eq:J}
\end{equation}
The proper time moment, $\tau^*$, is found from the condition $t - t(\tau^*) = 0$. 

The manifest form of the Maxwell equations in the background  (\ref{eq:metric}) with arbitrary profile $r(\rho)$ read
\bs\label{eq:Maxwell-Gen}
\begin{eqnarray}
-4\pi J^\rho &=& A^\rho_{,\rho\rho} + \frac{2r'}{r^2} A^\rho_{,\rho} - \frac{1}{r^2} \hat{L}^2 A^\rho - A^\rho_{,tt} - \frac{2r'^2}{r^2} A^\rho - \frac{2}{r} \left( \cot \theta A^\theta + A^\theta_{,\theta} + A^\varphi_{,\varphi}\right) - R^\rho_\rho A^\rho, \adb\\  
-4\pi J^\theta &=& A^\theta_{,\rho\rho} + \frac{4r'}{r} A^\theta_{,\rho} - \frac{1}{r^2} \hat{L}^2 A^\theta - A^\theta_{,tt} - \frac{1}{r^2} A^\theta \left( \cot^2\theta - r'^2 - r r'' \right) \adb\nonumber \\
&+& \frac{2}{r^3} \left( A^\rho_{,\theta} r' - r\cot \theta A^\varphi_{,\varphi} \right)  - R^\theta_\theta A^\theta , \adb\\  
-4\pi J^\varphi &=& A^\varphi_{,\rho\rho} + \frac{4r'}{r} A^\varphi_{,\rho} - \frac{1}{r^2} \hat{L}^2 A^\varphi - A^\varphi_{,tt} + \frac{2\cot\theta}{r^2}  A^\varphi_{,\theta} - \frac{1 - r'^2 - r r''}{r^2} A^\varphi \adb\nonumber \\
&+& \frac{2}{r^3} \left( \frac{r'}{\sin^2\theta }  A^\rho_{,\varphi} + \frac{r\cos\theta}{\sin^3\theta} A^\theta_{,\varphi} \right) - R^\varphi_\varphi A^\varphi , \adb\\ 
-4\pi J^t &=& A^t_{,\rho\rho} + \frac{2r'}{r^2} A^t_{,\rho} - \frac{1}{r^2} \hat{L}^2 A^t - A^t_{,tt} , 
\end{eqnarray}  
\es
where $\hat{L}^2 = \partial_\theta^2 + \cot\theta \partial_\theta + \csc^2\theta \partial_\theta^2$. Let us apply these equations for the case of the infinitely short throat  $r(\rho) = a + |\rho|$ ($r' = \textrm{sgn}(\rho),\ r'' = 2\delta (\rho)$). As usual in this case we consider the Maxwell equations out of the throat $\rho =0$ and additionally we find matching conditions on the throat by integrating the Maxwell equations over $\rho$ around the throat $(-\varepsilon,\varepsilon)$ followed by the limit $\varepsilon\to 0$. The potential $A^\mu$ remains to be continuous on sphere $\rho =0$. A discontinuous potential produces a discontinuous electromagnetic field which means emergence of charges and currents on the throat. Integrating as noted above we obtain the conditions
\bs\label{eq:Cond_at_zero}
\begin{eqnarray}
\left[ A^{i}_{,\rho} \right] &=& - \frac{4}{a} A^i(0),\adb \\
\left[ A^{t}_{,\rho} \right] &=& 0, 
\end{eqnarray}  
\es
where $\left[ f \right] = f(+0) - f(-0)$. Therefore, we have to set $A^i(0) =0$ and consider the Maxwell equation (\ref{eq:Maxwell-Gen}) out of the sphere $\rho =0$. In this case we obtain the Maxwell equations in two flat Minkowsky spacetimes, 
\begin{equation}\label{eq:Maxwell-Gen-Mink}
\square A^\mu = -4\pi J^\mu,
\end{equation} 
with matching conditions (\ref{eq:Cond_at_zero}) where $A^i(0) =0$ on this sphere. Therefore, the potential is $C^1$-regular on the throat and it has to vanish on the throat.  

To solve the Maxwell equations (\ref{eq:Maxwell-Gen-Mink}) we find the covariantly constant vectors field must obey the equations    
\begin{equation}
\nabla_\nu \os\pm\xi^a_\mu =0,
\end{equation}
for both parts of the wormhole spacetime with $\rho >0$ and $\rho <0$.  We mark over these parts by signs $\pm$. The solutions read  
\bs\label{eq:xi}
\begin{eqnarray}
\os\pm\xi^1_\mu &=& (\pm\sin\theta \sin\varphi, r(\rho) \cos\theta \sin\varphi, r(\rho) \sin\theta \cos\varphi, 0), \\
\os\pm\xi^2_\mu &=& (\pm\sin\theta \cos\varphi, r(\rho) \cos\theta \cos\varphi, -r(\rho) \sin\theta \sin\varphi, 0), \\
\os\pm\xi^3_\mu &=& (\pm\cos\theta, -r(\rho) \sin\theta, 0, 0), \\
\os\pm\xi^4_\mu &=& (0,0,0,1),
\end{eqnarray}
\es
with $\det(\os\pm\xi) = \sqrt{-g}$. Let us define four scalar functions $A^a = A^\mu \xi^a_\mu$ for $\rho\not =0$. These functions obey four scalar equations  
\begin{equation}
\square A^a = -4\pi J^a
\end{equation} 
out of the sphere $\rho =0$. The solution of these equations, $(dx' = \sqrt{-g(x')} d^4x')$
\begin{equation}\label{eq:A^a}
A^a(x) = \int G(x,x')J^a(x') dx',
\end{equation}
is expressed in terms of the scalar Green function which satisfies the equation,
\begin{equation}\label{eq:Green-Scalar-Gen}
\square G = - \delta(x,x'),
\end{equation}
with appropriate boundary conditions. 

The potential $A_\mu$ itself maybe found from Eq. (\ref{eq:A^a}) by the following expressions
\begin{equation}
A^\mu(x) = \int \xi^\mu_a(x)\xi_\nu^a(x')G(x,x')J^\nu(x')dx',\label{eq:Gmn}
\end{equation}
or in manifest form 
\begin{eqnarray}
A_i &=& \int GZ''_{,ik'}J^{k'}  dx',\adb\nonumber\\
A_4 &=& \int GJ^{4'} dx',
\end{eqnarray}
where we have used the twopoint function $Z = r(\rho)r(\rho')\cos\gamma$, and $\cos\gamma = \sin\theta\sin\theta'\cos (\varphi-\varphi') + \cos\theta \cos\theta'$. As a consequence of the conditions on the throat $A^i(0) =0$, we have to obey the relations $G(0,x') = G(x,0) =0$.  

To find the scalar Green function satisfying the Eq. (\ref{eq:Green-Scalar-Gen}) we solve the additional eigenvalue problem
\begin{equation}
\square \Psi = - \lambda^2 \Psi.
\end{equation}
The solution of the problem has the following form ($\lambda^2 = p^2 - \omega^2$)
\begin{equation}
\Psi_{p,\omega,l,m} = \frac{p}{\pi} e^{-i\omega t} Y_{lm}(\Omega) \Phi_p (\rho),
\end{equation}
where 
\begin{equation}
\Phi_p =  \left\{ 
\begin{array}{ll}
\phi_1 = k_1 j_l[p (a+\rho)] + k_2 y_l[p (a+\rho)], \rho>0,\\
\phi_2 = k_3 j_l[p (a-\rho)] + k_4 y_l[p (a-\rho)], \rho<0
\end{array} ,
\right.
\end{equation}
is the radial $C^1$-regular at the throat functions, and ($q=p a$)
\bs
\begin{eqnarray}
k_3 &=& +k_1 \left(2 q^2 y_l(q) j'_l(q)  + 1\right)+k_2 \left(2 q^2 y_l(q)    y'_l(q) \right), \adb\\
k_4 &=& -k_2 \left(2 q^2 y_l(q) j'_l(q)  + 1\right)- k_1 \left(2 q^2 j_l(q) j'_l(q)  \right).
\end{eqnarray}
\es
Here $j_l$ and $y_l$ are the spherical Bessel functions. This set of functions is orthogonal 
\begin{equation}
\int_{-\infty}^{+\infty} dt \int d\Omega \int_{-\infty}^{+\infty} r^2(\rho) d\rho \Psi_{p,\omega,l,m}(x)\Psi^*_{p',\omega',l',m'}(x) = \delta_{l,l'} \delta_{m,m'} \delta (\omega - \omega')  \delta (p - p'),
\end{equation}
due to the relation
\begin{equation}\label{eq:cond-1}
|k_1|^2 + |k_2|^2 + |k_3|^2 + |k_4|^2  = 1.
\end{equation}
Therefore we have $3$ conditions for $4$ constants. Now we demand the last condition that the vector potential should be zero at the throat $\rho =0$. This gives the relations
\begin{equation}
k_2 = -k_1  \frac{j_l(q)}{y_l(q)},\ k_3 = - k_1, \ k_4 = -k_2,
\end{equation}
and we obtain the following radial function
\begin{equation}
\Phi_p =  \left\{ 
\begin{array}{ll}
\phi_1 = +k_1 \left( j_l(p r) - \frac{j_l(q)}{y_l(q)} y_l(p r)\right), \rho>0 \\
\phi_2 = -k_1 \left( j_l(p r) - \frac{j_l(q)}{y_l(q)} y_l(p r)\right), \rho<0
\end{array}.
\right.
\end{equation}
The condition of orthogonality (\ref{eq:cond-1}) defines the last constant 
\begin{equation}
|k_1|^2 = \frac{1}{2}\frac{y_l^2(q)}{y_l^2(q) + j_l^2(q)}.
\end{equation}

The fullness condition, 
\begin{equation}
\sum_{lm}\int_{-\infty}^{+\infty} d\omega \int_{-\infty}^\infty d p   \Psi_{p,\omega,l,m}(x)\Psi^*_{p,\omega,l,m}(x') = \delta (t - t') \frac{\delta (\rho - \rho')}{r^2(\rho)}\frac{\delta (\theta - \theta') \delta (\varphi - \varphi')}{\sin \theta},\label{eq:Fullfill}
\end{equation}
is fulfilled with help of the relations $j_l(-x) = (-1)^lj_l(x), y_l(-x) = (-1)^{l+1}y_l(x)$. Indeed, integrating over $\omega$ we obtain the following expression
\begin{equation}
\delta (t - t') \sum_{lm}  Y_{lm}(\Omega)Y_{lm}^*(\Omega')\int_{0}^{\infty}  \frac{2}{\pi}p^2 Q_l(pr,pr',pa)dp,
\end{equation}
where 
\begin{eqnarray}
Q_l(u,u',q) &=&  \frac{\left\{j_l(u) y_l(q) - y_l(u) j_l(q)\right\}\left\{j_l(u') y_l(q) - y_l(u') j_l(q)\right\}}{y_l^2(q) + j_l^2(q)}\adb\nonumber\\
&=&  \frac{\pi}{2p\sqrt{rr'}}\frac{\left\{J_\nu(u) Y_\nu(q) - Y_\nu(u) J_\nu(q)\right\}\left\{J_\nu(u') Y_\nu(q) - Y_\nu(u') J_\nu(q)\right\}}{Y_\nu^2(q) + J_\nu^2(q)}.
\end{eqnarray}
To calculate the integral over $p$ we use the Weber-Orr transformation \cite[\S 7.10.5]{Bateman:1953:Htf-2}. This transformation leads to the relation
\begin{equation}
\int_0^\infty \frac{\left\{J_\nu(u) Y_\nu(q) - Y_\nu(u) J_\nu(q)\right\}\left\{J_\nu(u') Y_\nu(q) - Y_\nu(u') J_\nu(q)\right\}}{Y_\nu^2(q) + J_\nu^2(q)} p dp = \frac{\delta(r-r')}{r}.
\end{equation}
Therefore,
\begin{equation}
\int_{0}^{\infty} \frac{2}{\pi} p^2 Q_l(pr,pr',pa) dp =  \frac{\delta(\rho - \rho')}{r^2},
\end{equation}
and the relation (\ref{eq:Fullfill}) is fulfilled.

Taking into account these set of functions we obtain the retarded and advanced scalar Green functions in the following form 
\begin{equation}
G^{ret}_{adv}(x,x') = \frac{1}{\pi}\sum_{lm}Y_{lm}(\Omega)Y^*_{lm}(\Omega') \int_{-\infty}^\infty \frac{d\omega}{2\pi} \int_{0}^\infty \frac{p^2 dp e^{-i\omega (t-t')}}{p^2 -(\omega \pm i0)^2} Q_l(u,u',q),
\end{equation}
where 
\begin{equation}
Q_l(u,u',q) =  \frac{\left\{j_l(u) y_l(q) - y_l(u) j_l(q)\right\}\left\{j_l(u') y_l(q) - y_l(u') j_l(q)\right\}}{y_l^2(q) + j_l^2(q)},
\end{equation}
and $u=p(a+|\rho|), u'=p(a+|\rho'|), q= pa$. The radiative Green function,
\begin{equation}
G^{rad} = \frac{1}{2} \left(G^{ret} - G^{adv}\right),
\end{equation}
is expressed in the following form
\begin{equation}
G^{rad}(x,x') = \frac{i}{\pi}\sum_{lm}Y_{lm}(\Omega)Y^*_{lm}(\Omega') \int_{-\infty}^\infty \textrm{sgn}(\omega)d\omega \int_{0}^\infty p^2 dp e^{-i\omega (t-t')} \delta (p^2 - \omega^2)Q_l(u,u',q).\label{eq:Grad}
\end{equation}  

\section{Bremsstrahlung} \label{Sec:Brems}
The total energy radiated by a particle  throughout its lifetime reads
\begin{equation}
\mathcal{E} = \int T^\nu_{\mu;\nu} \xi^\mu dx,
\end{equation}
where $T_{\mu\nu}$ is the energy-momentum tensor of the electromagnetic field, and $\xi^\mu$ is the time-like Killing vector of the space-time. The space-time under consideration with metric (\ref{eq:metric}) admits the time-like Killing vector. Taking into account the electromagnetic energy-momentum tensor, we arrive at the following expression  
\begin{equation}
\mathcal{E} = 4\pi \iint G^{rad}_{\mu\nu',t} J^\mu J^{\nu'} dx dx'. \label{eq:E-main}
\end{equation}
This expression was obtained by Aliev \& Galtsov in Ref. \cite{Aliev:1989:Garsgcs} and was applied for the case of the cosmic string spacetime. Taking into account Eqs. (\ref{eq:J}), (\ref{eq:xi}), (\ref{eq:Gmn}) and (\ref{eq:Grad}) we arrive at the following formula for the total energy loss
\begin{equation}
\mathcal{E} = 2\pi e^2 \sum_{lm} \int_{0}^{\infty} \frac{d\omega \omega M^{(a)}_{lm}M_{(a)lm}}{J_\nu^2(\omega a) +Y_\nu^2(\omega a)} ,
\end{equation}
where
\begin{equation}
M^{(a)}_{lm} = \int_{-\infty}^{+\infty}dt e^{i\omega t}\frac{v^{a}(t) Y_{lm}(\theta (t), \varphi (t))}{\sqrt{r(t)}} \left\{J_\nu(\omega r(t))Y_\nu(\omega a) - Y_\nu(\omega r(t)) J_\nu(\omega a)\right\},
\end{equation}
and $v^{a}(t) = \xi^a_\mu (t) v^\mu(t)$. Here $(r(t) = a + |\rho (t)|,\theta (t),\varphi (t))$ is a trajectory of the particle. Also we define the spectral density of the energy loss by the relation 
\begin{equation}
\mathcal{E}(\omega) = 2\pi e^2 \sum_{lm} \frac{\omega M^{(a)}_{lm}M_{(a)lm}}{J_\nu^2(\omega a) +Y_\nu^2(\omega a)}.
\end{equation}

Let us apply this expression for a uniformly moving particle in the background of a wormhole with an infinitely short throat. The trajectory of the particle moving in one part of the wormhole spacetime is described by the relations
\begin{equation}
r(t) = a + \rho(t)= \sqrt{b^2 + v^2t^2},\ \varphi = \arctan \frac{vt}{b},\ \theta = \frac{\pi}{2},\ u^t = \gamma = \frac{1}{\sqrt{1-v^2}}.
\end{equation}
Here $b\geq a$ -- impact parameter. This is closest distance to wormhole's throat and we set at the moment $t=0$ and angle $\varphi = 0$. The $3$-velocity components read: $v^\rho = v^2t/r(t),\ v^\varphi = vb/r^2(t),\ v^\theta =0,\ v^t =1$.  For this trajectory $v^a(t) v_a(t') = v^2 -1$ and the total energy reads ($\nu = l+1/2$)
\begin{equation}
\mathcal{E} = \frac{2\pi e^2}{\gamma^2} \sum_{lm}\left|Y_{lm}(\frac{\pi}{2},0)\right|^2 \int_{0}^{\infty} \frac{d\omega \omega |N^m_l(\omega)|^2}{J_\nu^2(\omega a) +Y_\nu^2(\omega a)} ,
\end{equation}
where
\begin{eqnarray}
N^m_l(\omega) &=& \int_{-\infty}^{+\infty}dt \frac{e^{i\omega t -im \varphi (t)}}{\sqrt{r(t)}} \left\{J_\nu(\omega r(t))Y_\nu(\omega a) - Y_\nu(\omega r(t)) J_\nu(\omega a)\right\}\adb\nonumber\\
&=& \frac{\sqrt{b}}{v} \left\{ \mathcal{J}_l^m(\omega)Y_\nu(\omega a) - \mathcal{Y}_l^m(\omega)J_\nu(\omega a)\right\},
\end{eqnarray}
and ($t=bx/v$)
\bs
\begin{eqnarray}
\mathcal{J}_l^m(\omega) &=&\int_{-\infty}^{+\infty}dx e^{i\frac{\omega b}{v}x -im \arctan x }\frac{J_\nu (\omega b\sqrt{1+x^2})}{\sqrt[4]{1+x^2}},\adb \label{eq:J^m_l}\\
\mathcal{Y}_l^m(\omega) &=& \int_{-\infty}^{+\infty}dx e^{i\frac{\omega b}{v}x -im \arctan x }\frac{Y_\nu (\omega b\sqrt{1+x^2})}{\sqrt[4]{1+x^2}}.\label{eq:Y^m_l}
\end{eqnarray}
\es
The spherical functions above,
\begin{equation}
Y_{lm}(\frac{\pi}{2},0) = \sqrt{\frac{2l+1}{4\pi}} \sqrt{\frac{(l-m)!}{(l+m)!}} P_l^m (0),
\end{equation}
equal zero if $l+m=2k+1$ -- odd number, because $P_l^m(0) = 0$ in this case. For this reason the sum, $l+m=2k$, has to be an even number. 

Let us consider the first expression (\ref{eq:J^m_l}) as an integral over  the complex plane of $x$. There are two branch points $x=\pm i$ and two cuts, $(-i\infty, -i)$ and $(i,i\infty)$. We shift the contour of integration to the upper half-plane and put it to the left and right banks of cut and small circle around branch point $x=i$. The integral over branch point equals zero and two integrals over banks gives the following expression:
\begin{equation}
\mathcal{J}_l^m(\omega,p) = -2  \sin\frac{\pi}{2} \left(\nu - m -\frac{1}{2}\right) \int_1^\infty dy e^{-\frac{\omega b}{v}y} \left(\frac{y-1}{y+1}\right)^{-\frac{m}{2}} \frac{I_\nu (pb\sqrt{y^2-1})}{\sqrt[4]{y^2-1}},
\end{equation}
where $I_\nu$ is the modified Bessel function. As noted above, the sum $l+m=2k$, then $\nu - m -\frac{1}{2} = 2(k-m)$ and $\mathcal{J}_l^m(\omega,p) =0$ because  $\sin\frac{\pi}{2} \left(\nu - m -\frac{1}{2}\right) = \sin \pi (k-m)=0$. The same approach can not be applied for the integral (\ref{eq:Y^m_l}). Indeed, the integral the over circle is divergent because the Neumann function has a singularity for zero argument and we can not obtain a finite expression.  Therefore we arrive at the following expression:
\begin{equation}
\mathcal{E} = \frac{2\pi e^2 b}{v^2 \gamma^2} \sum_{lm}\left|Y_{lm}(\frac{\pi}{2},0)\right|^2 \int_{0}^{\infty} \frac{d\omega \omega J_\nu^2(\omega a)}{J_\nu^2(\omega a) +Y_\nu^2(\omega a)} |\mathcal{Y}_l^m(\omega)|^2,\label{eq:Eb}
\end{equation}
where
\begin{equation}
\mathcal{Y}_l^m(\omega) =  \int_{-\infty}^{+\infty}dx e^{i\frac{\omega b}{v}x -im \arctan x }\frac{Y_\nu (\omega b\sqrt{1+x^2})}{\sqrt[4]{1+x^2}}.\label{eq:Ymain}
\end{equation}

Then  the spectral density reads
\begin{equation}
\mathcal{E}(\omega) = \frac{2\pi e^2 b}{v^2 \gamma^2} \sum_{lm}\left|Y_{lm}(\frac{\pi}{2},0)\right|^2  \frac{\omega J_\nu^2(\omega a)|\mathcal{Y}_l^m(\omega)|^2}{J_\nu^2(\omega a) +Y_\nu^2(\omega a)}.\label{eq:E(w)}
\end{equation}

We are coming now to numerical calculations of the spectral density of the energy (\ref{eq:E(w)}) with its preliminary analysis in some limiting cases. Let us define the new variable $\sigma = \frac{\omega b}{v\gamma}$ and new function $\bar{\mathcal{E}}(\sigma) = \frac{b^2}{e^2a^2} \mathcal{E}(\sigma \frac{v\gamma}{b})$. Then the full energy has the following form
\begin{equation}
\mathcal{E} =  \frac{e^2a^2}{b^3} v\gamma\int_0^\infty \bar{\mathcal{E}}(\sigma) d\sigma.\label{eq:Ebar}
\end{equation}

In the case of nonrelativistic particles, $v \ll 1$, the main contribution comes from the term with $l=m=0$. For these numbers \cite[2.4.16(3)]{Prudnikov:1981:VISEf}
\begin{equation}
\mathcal{Y}_0^0(\omega) = - \sqrt{\frac{8}{\pi\omega b}} K_0 \left(\frac{\omega b}{v\gamma}\right),
\end{equation}
 and we obtain the following expression for the spectral density
\begin{equation}
\bar{\mathcal{E}}(\sigma) \approx \frac{4}{\pi} \sigma^2 K_0^2(\sigma), 
\end{equation}
where $K_0(\sigma)$ is the modified Bessel function. This expression does not depend on the velocity, $v$, and impact factor $b$. The spectrum falls down exponentially fast, $\bar{\mathcal{E}}(\sigma) \approx 2\sigma e^{-2\sigma}$ and tends to zero as $4\sigma^2 \ln^2\sigma/\pi$ at the origin. Integrating over $\sigma$ we arrive at the following expression for the total energy loss in the nonrelativistic case
\begin{equation}
\mathcal{E} \approx e^2 \frac{\pi va^2}{8b^3}. 
\end{equation}  
Therefore, the total energy is proportional to the first power of the velocity. In the case of the cosmic string \cite{Aliev:1989:Garsgcs} and global monopole \cite{Bezerra:2002:Bgfgm} spacetimes the energy is proportional to the third degree of the velocity. The spectral density of the energy has a maximum at the point $\sigma \approx 3/5$, that is at the frequency 
\begin{equation}
\omega_m = \frac{3\gamma v}{5b}. 
\end{equation}

Let us consider the ultrarelativistic case, $v\to 1$. First of all we extract divergent (in this limit) part in the integral (\ref{eq:Ymain}). It is easy to see that 
\begin{equation} \label{eq:Yexp}
\frac{Y_{l+1/2}(t)}{\sqrt{t}} = 
\left\{ 
\begin{array}{ll}
P_n \frac{\cos t}{t} + Q_{n-1} \frac{\sin t}{t^2}, & l =2n \\
M_n \frac{\cos t}{t^2} + N_{n} \frac{\sin t}{t}, & l =2n+1
\end{array},
\right.
\end{equation}
where $P_n,Q_n,M_n,N_n$ are polynomials of $n$-th order of the variable $1/t^2$. The integrand (\ref{eq:Ymain}) incorporates this expression with $t = \omega b \sqrt{1+x^2}$.  The divergence for $v\to 1$ appears only for lowest power $t$ in denominator in Eq. (\ref{eq:Yexp}), that is for 
\begin{equation} 
\left.\frac{Y_{l+1/2}(t)}{\sqrt{t}}\right|_{div} = -\sqrt{\frac{2}{\pi}}\frac{(-1)^n}{t}
\left\{ 
\begin{array}{ll}
\cos t, & l =2n \\
\sin t, & l =2n+1
\end{array}.
\right.
\end{equation}
First of all let us consider the case $m=0$. The divergent part for $v\to 1$ reads
\begin{equation}
\left.\mathcal{Y}_l^0(\omega)\right|_{div} = -   \sqrt{\frac{2}{\pi \omega b}}(-1)^n \int_{-\infty}^{+\infty} \frac{e^{i \frac{\omega b}{v}x}}{\sqrt{1+x^2}} \left\{ \genfrac{}{}{0 pt}{}{\cos (\omega b \sqrt{1+x^2})}{\sin (\omega b \sqrt{1+x^2})} \right\}dx,
\end{equation}  
where the upper case is for $l=2n$ and down case is for $l=2n+1$. By using Eq. 2.5.25(15) from textbook \cite{Prudnikov:1981:VISEf} we obtain
\begin{equation}
\left.\mathcal{Y}_{2n}^0(\omega)\right|_{div} = -   \sqrt{\frac{8}{\pi \omega b}}(-1)^n K_0 \left(\frac{\omega b}{v\gamma}\right),
\end{equation}
which is zero for $l=2n+1$ (see Eq. 2.5.25(9) in Ref. \cite{Prudnikov:1981:VISEf}).   

For the case $m\not = 0$ we have an additional factor $e^{im \frac{\pi}{2}\textrm{sgn}(x)}$ in the integrand. Because the sum $l+m$ must be an even number, the  parities of $m$ and $l$ should be the same. For even $l=2n$ and $m=2j$ we obtain
\begin{equation}
\left.\mathcal{Y}_{2n}^{2j}(\omega)\right|_{div}  = -   \sqrt{\frac{8}{\pi \omega b}}(-1)^n \int_{0}^{\infty} \cos \left(\frac{\omega b}{v}x\right) \frac{ \cos (\omega b \sqrt{1+x^2})}{\sqrt{1+x^2}}dx =  -   \sqrt{\frac{8}{\pi \omega b}}(-1)^{n+j} K_0 \left(\frac{\omega b}{v\gamma}\right).
 \end{equation}
 
For odd $l=2n+1$ and $m=2j+1$ the integral,   
\begin{equation}
\left.\mathcal{Y}_{2n+1}^{2j+1}(\omega)\right|_{div} =   \sqrt{\frac{8}{\pi \omega b}}(-1)^{n+j} \int_{0}^{\infty} \sin \left(\frac{\omega b}{v}x\right) \frac{ \sin (\omega b \sqrt{1+x^2})}{\sqrt{1+x^2}}dx,
 \end{equation}
can not be found in manifest form. Nevertheless we may extract the divergent part of this integral:
\begin{equation}
\left.\mathcal{Y}_{2n+1}^{2j+1}(\omega)\right|_{div} =   \sqrt{\frac{8}{\pi \omega b}}(-1)^{n+j} K_0 \left(\frac{\omega b}{v\gamma}\right) - \sqrt{\frac{8}{\pi \omega b}}(-1)^{n+j} \int_{0}^{\infty}  \frac{ \cos \left(\frac{\omega b}{v}x + \omega b \sqrt{1+x^2}\right)}{\sqrt{1+x^2}}dx.
 \end{equation}   
The last integral is finite even for $v=1$.  Therefore, in the limit $v\to 1$ for all cases we obtain the main contribution
\begin{equation}
|\mathcal{Y}_{l}^{m}(\omega)|^2 \approx  \frac{8}{\pi \omega b} K_0^2 \left(\frac{\omega b}{v\gamma}\right),
\end{equation}
and the spectral density reads
\begin{equation}
\mathcal{E}(\omega) \approx \frac{16 e^2}{v^2 \gamma^2} K_0^2 \left(\frac{\omega b}{v\gamma}\right)\sum_{lm}\left|Y_{lm}(\frac{\pi}{2},0)\right|^2  \frac{J_\nu^2(\omega a)}{J_\nu^2(\omega a) +Y_\nu^2(\omega a)}.
\end{equation}
The numerical analysis of the function 
\begin{equation}
f(x) = \sum_{lm}\left|Y_{lm}(\frac{\pi}{2},0)\right|^2  \frac{J_\nu^2(x)}{J_\nu^2(x) +Y_\nu^2(x)} = \sum_{l=0}^\infty\frac{2l+1}{4\pi} \frac{J_\nu^2(x)}{J_\nu^2(x) +Y_\nu^2(x)}
\end{equation}
with a subsequent Pade approximation shows that 
\begin{equation}
  f(x) = \frac{1}{16\pi}(2.4 x +  x^2).
  \end{equation}

Taking into account this expression we obtain the spectral density 
\begin{equation}
\mathcal{E}(\omega) \approx \frac{e^2 \omega a}{\pi v^2 \gamma^2} K_0^2 \left(\frac{\omega b}{v\gamma}\right)(2.4 +  \omega a),
\end{equation}
and 
\begin{equation}
\bar{\mathcal{E}}(\sigma) \approx \frac{\sigma}{\pi}  K_0^2 (\sigma)\left( \frac{2.4 b}{av\gamma} + \sigma \right).
\end{equation}
In the case of $\gamma \gg b/a$ we arrive at the simple formula
\begin{equation}
\bar{\mathcal{E}}(\sigma) \approx \frac{\sigma^2}{\pi}  K_0^2 (\sigma).
\end{equation}
The spectrum falls down exponentially fast, $\bar{\mathcal{E}}(\sigma) \approx \sigma e^{-2\sigma}/4$ and tends to zero as $\sigma^2\ln^2\sigma /\pi$ at the origin. Integrating this function (see Eq. (\ref{eq:Ebar})) we obtain the total energy in the ultrarelativistic case
\begin{equation}
 \mathcal{E} \approx e^2 \frac{\pi \gamma a^2}{32 b^3}, 
 \end{equation}
with characteristic frequency 
 \begin{equation}
\omega_m = \frac{3\gamma v}{5b}. 
\end{equation}     

\begin{figure}[ht]
\includegraphics[width=8cm]{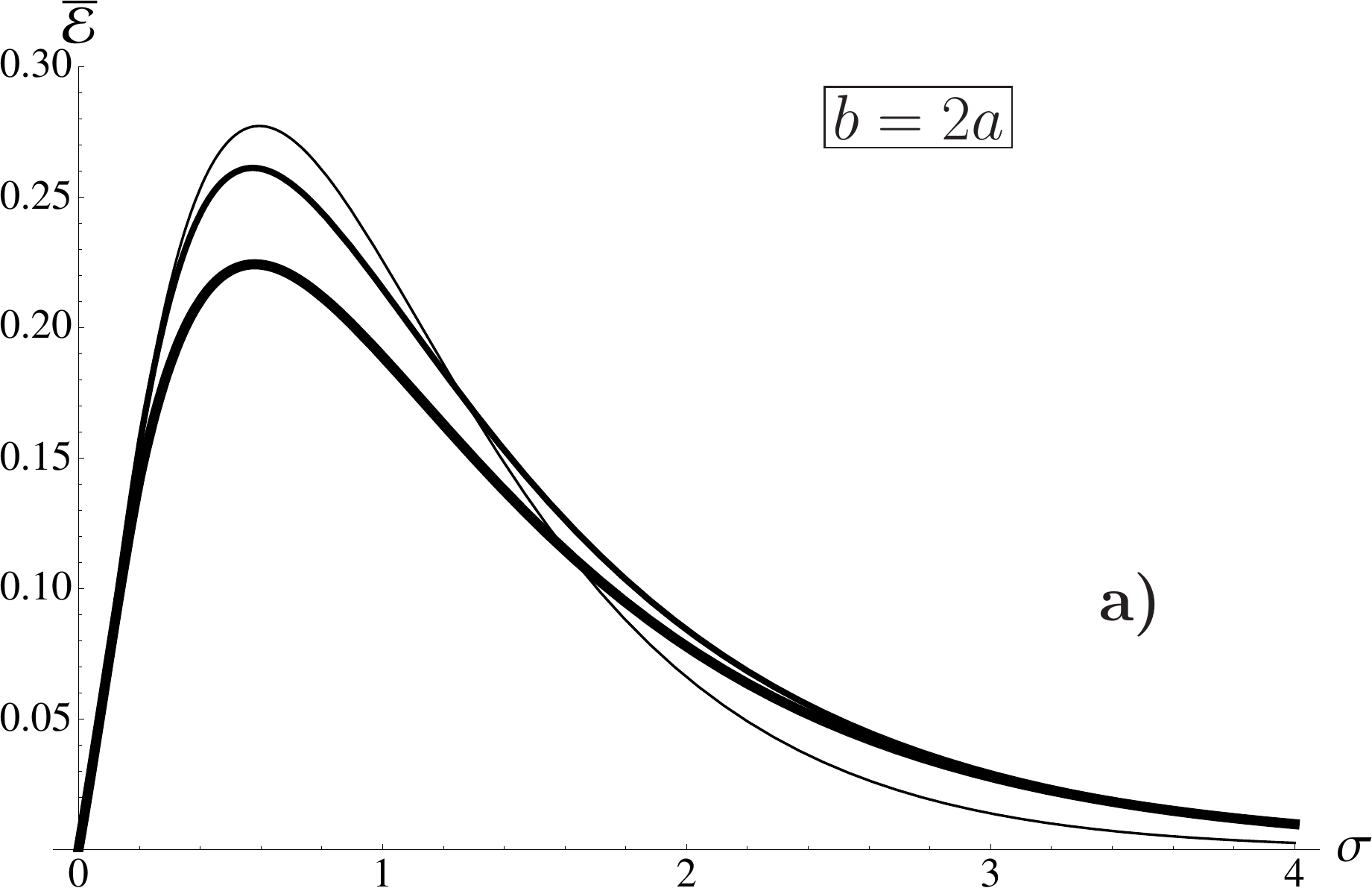}\includegraphics[width=8cm]{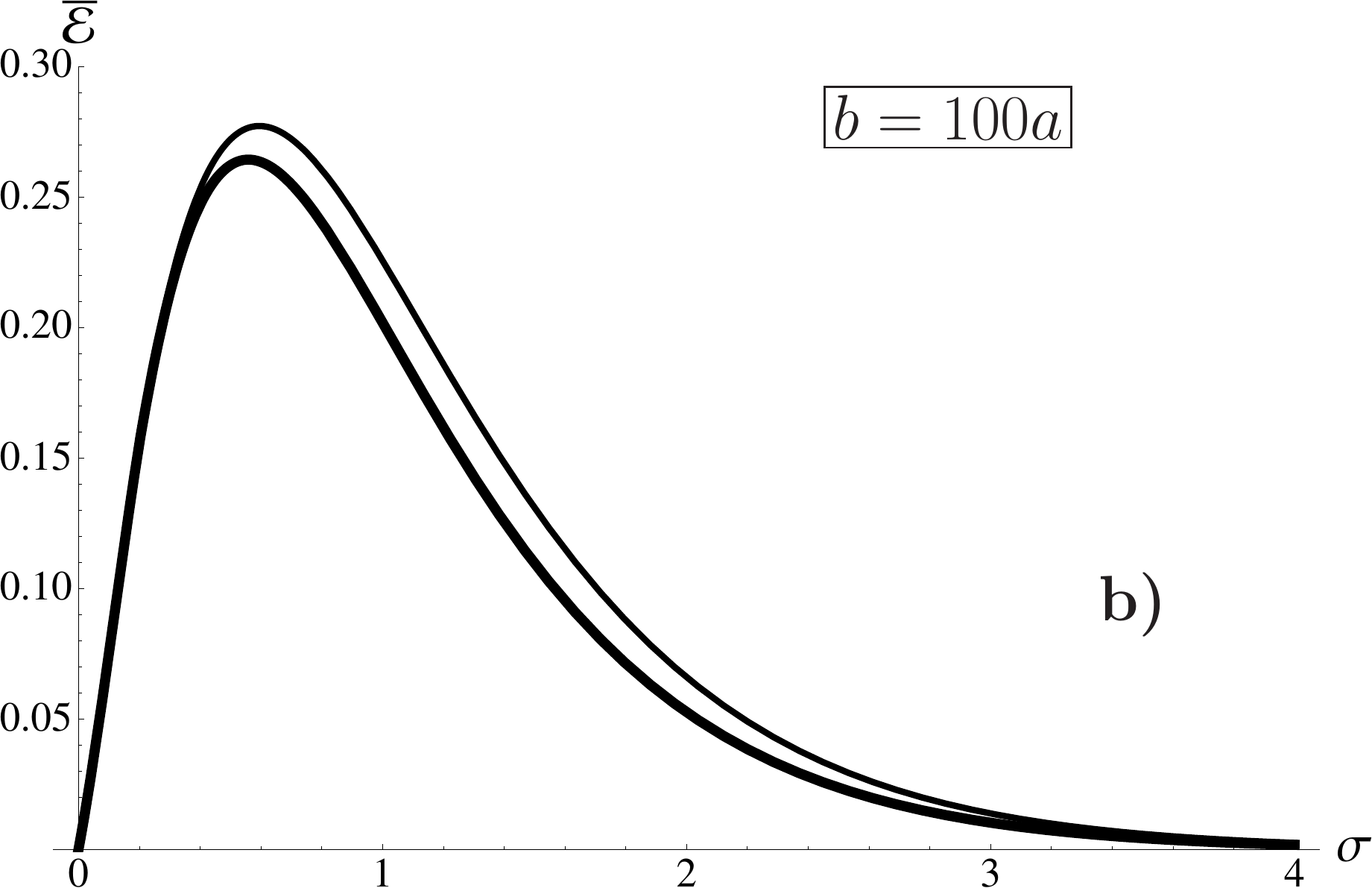}
\caption{The spectral density of the energy (\ref{eq:E(w)}) as a function of $\sigma = \frac{\omega b}{v\gamma}$: (a) $v=0$ (thin line), $v=0.9$ (middle thickness line) and $v=0.99$ (thick line) (b) The thin line is for $v=0$ up to $v=0.99$ and the thick line is for $v=0.999$. The extrema of the energy are located at  $\sigma \sim 1$, that is for frequency $\omega \sim v\gamma/b$.}\label{fig:esigma}
\end{figure}

The numerical  evaluations of  $\bar{\mathcal{E}}(\sigma)$ are shown in Fig. \ref{fig:esigma} for different values of velocity and impact factors $b/ a=2,100$.  We observe agreement between theoretical and numerical considerations. 

Let us consider a particle which moves through the wormholes throat by radial trajectory $r(t) = a + v |t|, \varphi =0, \theta = \pi/2$ with velocity $v^\rho = v, v^\theta = v^\varphi =0$. Positive and negative time corresponds to different parts of the wormhole spacetime.  In this case $v^a(t) v_a (t') = v^2 \textrm{sgn} (t) \textrm{sgn} (t') -1$ and the spectral density of energy reads
\begin{eqnarray}
\mathcal{E}(\omega) &=& \frac{2e^2}{a v^2} \sum_{l=0}^\infty \frac{ (2l+1)\bar \omega}{J_\nu^2(\bar\omega) + Y_\nu^2(\bar\omega)} \nonumber \adb \\
&\times&\left\{ \left|\int_0^\infty \frac{dx \cos \frac{\bar\omega x}{v}}{\sqrt{1+x}} \left[J_\nu(\bar\omega (1+x)) Y_\nu (\bar\omega) - Y_\nu(\bar\omega (1+x)) J_\nu (\bar\omega )\right]\right|^2\right. \nonumber \adb \\
&-& \left. v^2 \left|\int_0^\infty \frac{dx \sin \frac{\bar\omega x}{v}}{\sqrt{1+x}} \left[J_\nu(\bar\omega (1+x)) Y_\nu (\bar\omega) - Y_\nu(\bar\omega (1+x)) J_\nu (\bar\omega )\right]\right|^2 \right\},\label{eq:energy-radial}
 \end{eqnarray} 
where $\bar\omega = \omega a$ is a dimensionless frequency.  

\begin{figure}[ht]
\includegraphics[width=8cm]{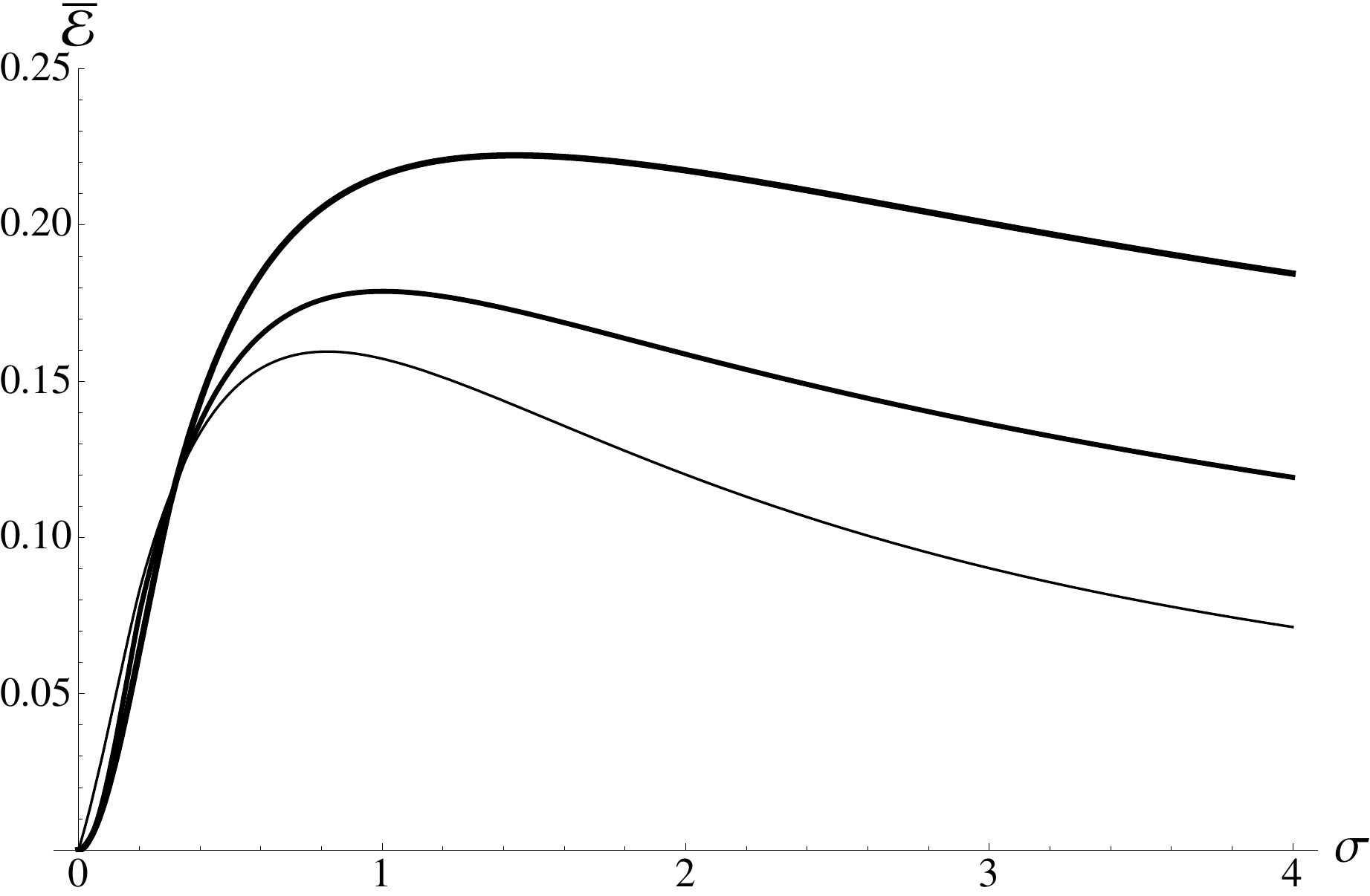}
\caption{The spectral density of the energy loss (\ref{eq:energy-radial}) for $v=0$ (thin line), $v=0.1$ (middle thickness line) and $v=0.3$ (thick line) as a function of $\sigma = \frac{\omega a}{v\gamma}$. The extrema are located at $\sigma \sim 1$, that is for $\omega \sim v\gamma/a$.}\label{fig:esigma-radial}
\end{figure}

Let us define in this case the new variable $\sigma = \frac{\omega a}{v\gamma}$, and new spectral density $\bar{\mathcal{E}}(\sigma) = \frac{a}{e^2} \mathcal{E}(\sigma \frac{v\gamma}{a})$. The full energy reads
\begin{equation}
\mathcal{E} = \frac{e^2}{a}v\gamma \int_0^\infty \bar{\mathcal{E}}(\sigma) d\sigma.
\end{equation}
In the nonrelativistic case, $v\ll 1$, the energy density has the following form
\begin{equation}
\bar{\mathcal{E}}(\sigma) = \frac{4}{\pi}\sigma^2 \left|\int_0^\infty \frac{\sin x dx}{(x+\sigma)^2} \right|^2,
\end{equation}
and the total energy $\mathcal{E} = \frac{2v}{3a}e^2$. The integral above may be expressed in terms of integral sine and cosine special functions
\begin{equation}
\int_0^\infty \frac{\sin x dx}{(x+\sigma)^2} = -\cos \sigma\ \mathrm{Ci}(\sigma) + \sin\sigma \left(\frac{\pi}{2} - \mathrm{Si}(\sigma)\right). 
\end{equation}
Taking into account this expression we find  maximum of spectral density at point $\sigma \approx 0.77$, that is for 
\begin{equation}
\omega_m = 0.77\frac{\gamma v}{a}. 
\end{equation}
The spectral density falls down slowly, $\bar{\mathcal{E}} \approx 4/\pi \sigma^2$ for $\sigma\to \infty$. In Fig. \ref{fig:esigma-radial} we show numerical simulations of the energy density $\bar{\mathcal{E}}(\sigma)$ for different values of the velocity. 

\section{Discussion and conclusion}\label{Sec:DC}

Let us here summarize the results obtained in the above sections. We calculated the total energy loss and spectral density for uniformly moving electrically charged particles in a  wormhole spacetime with an infinitely short throat. If a particle moves uniformly with impact parameter $b$ the total energy is given by Eq. (\ref{eq:Eb}). It is better to analyze the energy as a function of the new dimensionless variable $\sigma = \omega b/v\gamma$:
\begin{equation}
\mathcal{E} =  \frac{e^2 a^2}{b^3} v\gamma\int_0^\infty \bar{\mathcal{E}}(\sigma) d\sigma.
\end{equation}
In the nonrelativistic case, $v\ll 1$, the spectral density $\bar{\mathcal{E}} \sim \sigma^2 K_0^2(\sigma)$. The density exponentially falls down, $\bar{\mathcal{E}} \sim \sigma e^{-2\sigma}$ for large frequency $\sigma\to \infty$ and tends to zero, $\bar{\mathcal{E}} \sim  \sigma^2\ln^2\sigma$ for small frequencies, $\sigma \to 0$. The maximum of the density is for $\sigma \sim 1$, that is for $\omega \sim v/b$ and the total energy loss $\mathcal{E} \sim  v a^2/b^3$. In the opposite case of ultrarelativistic motion we have the same behavior of the spectral density for small and large $\sigma$. The total energy loss now is  $\mathcal{E} \sim \gamma a^2/b^3$ with a maximum at the point $\omega \sim \gamma/b$. We may combine both cases by statement that the total energy loss $\mathcal{E} \sim e^2 v \gamma a^2/b^3$ with a maximum at $\omega \sim v\gamma/b$. The spectrum exponentially falls down for large  $\sigma = \omega b/v\gamma$ and tends to zero at the origin. The typical plots of the spectral densities are shown in Fig.  \ref{fig:esigma}. 

For particles which move radially through the wormhole throat we obtain a similar picture. The spectral density has the form given by Eq. (\ref{eq:energy-radial}) and $\mathcal{E} \sim v\gamma/a$. The spectral energy has maximum at frequency $\omega \sim v\gamma/a$. In this case the energy falls down more slowly $\mathcal{E} \sim 1/\sigma^2$, where $\sigma = \omega a/v\gamma$. A typical spectral density of the energy is shown in Fig. \ref{fig:esigma-radial}. 

The wormhole spacetime considered here is everywhere flat except sphere with singular curvature. We observe that even uniformly moving particles radiate an electromagnetic field and the spectral density of the energy loss has a maximum at a specific frequency. In Minkowsky spacetime this effect is forbidden due to the energy conservation law. 

It is also important  to compare the above results with radiation in a different spherically symmetric background. This is  interesting in context of wormhole's mimicry \cite{Damour:2007:Wabhf}. It was claimed that all classical phenomena in a static spherically symmetric wormhole spacetime are the same as in an appropriate black hole spacetime and the  only way to distinguish them is through the observation of Hawking radiation. We already noted in Ref. \cite{Khusnutdinov:2007:Scpwst} that this is not the case for a self-interaction force, which has an opposite sign in these two geometries.  The origin may be connected with the fact that the self-force is obtained by a renormalization procedure and for this reason it is, in fact, a quantum phenomenon. 

In the geometry under consideration the trajectories of uniformly moving particles are straight lines. This is not the case for a Schwarzschild geometry and the main interest connects with close orbits of particle around central mass. In this case there is a well-known flat space-time synchrotron radiation. The synchrotron  radiation of a particle moving along a geodesic in the black hole space time was considered in detail in many papers. Let us refer here for papers \cite{Breuer:1973:VaTRfSRCG,Breuer:1973:PoSRfRSCG} and book \cite{Breuer:1975:Gptasr}. In these papers the spectral density of radiation was calculated for scalar ($s=0$), electromagnetic ($s=1$) and gravitational ($s=2$) fields in the background of a black hole. The main result is that the power spectrum has the following form 
\begin{equation}
\frac{d\mathcal{E}_s}{d\omega} \sim \left(\frac{\omega}{\omega_c}\right)^{1-s} e^{-\frac{2\omega}{\omega_c}},
\end{equation}
where $\omega_c$ is some critical frequency. We observe from this formula that in the case of the electromagnetic field ($s=1$) the spectrum is exponentially steadily decreasing without a characteristic frequency. Therefore we expect great difference in the radiation of plasma surrounding a black hole or a wormhole. The difference may be explained by a non-trivial topology of the wormhole's space-time. We have two copies of Minkowsky space-time which are glued together on the sphere. Some part of the electromagnetic field may go through the wormhole's throat to a second universe \cite{Khusnutdinov:2007:Scpwst}. 

The radiation of uniformly moving particles in the background of cosmic strings was considered in Ref.  \cite{Aliev:1989:Garsgcs} and in the background of a global monopole in Ref.  \cite{Bezerra:2002:Bgfgm}. The space-time of an infinitely  thin cosmic string  has cylindrical symmetry and is everywhere flat except at the origin where it has infinite curvature. The space-time of a point-like global monopole has spherical symmetry with nonzero curvature. In both cases the spectrum of radiation has no specific frequency and steadily falls down.  

Let us now speculate about result obtained. A particle moving near or through a wormhole will radiate electromagnetic waves. The magnitude and characteristic frequency of the radiation of each particle depends on the velocity and impact parameter of the particle's trajectory. If a particle moves through wormhole, the characteristic frequency depends on the  radius of the throat only. Therefore, we observe specific radiation from the plasma surrounding the wormhole's throat with a specific frequency. To obtain this frequency we should  average the energy over specific particles distribution and then find the maximum of the spectral density obtained. These calculations will be considered in separate paper. 

\section{Acknowledgments}
The author is grateful Fulbright Visiting Scholar Program for financial support and Department of Physics of University of South Florida, USA for hospitality. This work was supported in part by the Russian Foundation for Basic Research Grant No. 11-02-01162-a and No. 13-02-00757-a.



\end{document}